\documentclass[pre,aps,floats,superscriptaddress,floatfix,twocolumn,longbibliography]{revtex4-1}
\usepackage{amssymb,amsmath}
\usepackage{amsmath,amssymb}
\usepackage{enumitem,kantlipsum}
\usepackage{lipsum}
\usepackage{soul}
\usepackage{cancel}
\usepackage{gensymb}
\usepackage{graphicx}
\usepackage{psfrag}
\usepackage{color}
\usepackage{dcolumn}
\usepackage{bm}
\usepackage[normalem]{ulem}
\usepackage{comment}
\usepackage{hyperref}
\hypersetup{
  colorlinks=true,
  citecolor=magenta,
  linkcolor=cyan,
}

\newcommand{\gap}{\quad \quad \quad \quad \quad}
\begin{document}
\title{Strong-coupling phases of conserved growth models are crumpled}
\author{Debayan Jana}\email{debayanjana96@gmail.com}
\affiliation{Theoretical Physics Division, Saha Institute of Nuclear Physics, a CI of Homi Bhabha National Institute, 1/AF, Bidhannagar, Calcutta 700064, West Bengal, India}
\author{Abhik Basu}\email{abhik.123@gmail.com, abhik.basu@saha.ac.in}
\affiliation{Theoretical Physics Division, Saha Institute of Nuclear Physics, a CI of Homi Bhabha National Institute, 1/AF, Bidhannagar, Calcutta 700064, West Bengal, India}

\begin{abstract}
  We show that stochastically driven nonequilibrium conserved growth models admit generic strong coupling phases for sufficiently strong nonlocal chemical potentials underlying the dynamics. The models exhibit generic roughening transitions between perturbatively accessible weak coupling phases satisfying an exact relation between the scaling exponents in all dimensions $d$, and strong coupling phases. In dimensions at or below the critical dimension $d_c$, the latter phases are unstable and argued to be crumpled, and thus distinct from the well-known strong coupling rough phase of the Kardar-Parisi-Zhang equation in dimensions $d\geq 2$. At $d_c$, conventional spatiotemporal scaling in the weak coupling phase is logarithmically modulated and are exactly obtained. 
\end{abstract}

\maketitle

Strong coupling phases and roughening transitions in nonequilibrium systems have been an intense focus of research. 
 Kardar-Parisi-Zhang (KPZ) equation for surface growth without any conservation laws~\cite{kpz,stanley} undergoes a roughening transition from a smooth to a perturbatively inaccessible strong coupling rough surface at dimensions $d>2$, as the noise strength or the nonlinear coupling strength is increased; see Ref.~\cite{halpin_1} for a comprehensive review. Perturbative inaccessibility of the strong coupling regime of the KPZ equation has promoted researchers to use extensive numerical methods~\cite{kpz_num_1,kpz_num_2,kpz_num_3,kpz_num_4,kpz_num_5,kpz_num_6,kpz_num_7,kpz_num_8,kpz_num_9,kpz_num_10,kpz_num_11,kpz_cocktail,3d_kpz_num_1,3d_kpz_num_2,3d_kpz_num_3} for $d \ge 2$ or mode coupling theories (MCT)~\cite{jkb-mct,mct_cates,mct_frey,mct_moore,mct_tu,jkb-book,bmhd1,bmhd2} to extract the scaling behavior in the rough phase. Such roughening transitions to strong coupling phases have been observed in various generalizations of the KPZ equation, e.g., coupled KPZ~\cite{bmhd2,bmhd1} and KPZ with nonlocality~\cite{debayan1,debayan2}. Subsequently, to account for conserved growth processes, conserved versions of the KPZ equation were proposed and studied. This includes the conserved KPZ (CKPZ) equation~\cite{ckpz}, which belongs to a distinct universality class without any roughening transition. A related model~\cite{lds} (hereafter LDS equation), introduced to describe idealized molecular beam
epitaxy (MBE)~\cite{stanley,krug,lds}, has the same form as the CKPZ equation, but driven by nonconserved noises~\cite{stanley,lds}. 
The LDS equation belongs to a different universality class known as the LDS class that also lacks any roughening transition. See also Ref.~\cite{cuerno_conserved} for a partly related one-dimensional model study. 

The CKPZ or LDS models were subsequently generalized to the CKPZ+ model~\cite{ckpz_plus} and MBE+ model~\cite{mbe_plus} respectively by considering nonlocal contributions to the underlying chemical potential $\mu$. Both CKPZ+ and MBE+ models display roughening transitions, physically distinct from their counterpart in the KPZ equation~\cite{ckpz_plus,mbe_plus}. 

\begin{figure*}[]
\includegraphics[width=0.9\textwidth]{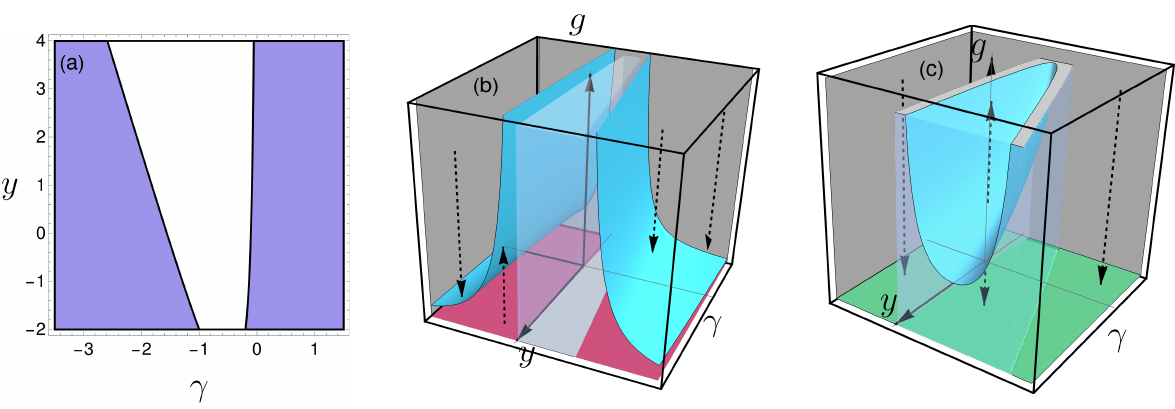}
\caption{($\bar\nu=0$) (a) Stable (violet) and unstable (white) regions in the $\gamma$-$y$ plane as obtained from RG and MCT calculations at $d =d_c= 4 + y$.~(b) RG flow diagram at $d < d_c$. The blue planes represent stable fixed planes toward which the RG flow lines (black dashed arrows) converge. The red region denotes the stable phase accessible via perturbation theory, while the central white region is speculated to be unstable.~(c) RG flow diagram for $d > d_c$. The central white region contains a blue plane representing an unstable fixed plane, where a roughening transition occurs between a smooth phase (green region) and a perturbatively inaccessible strong coupling rough phase, as indicated by the RG flow lines. Outside this region, the roughening transition disappears, and only the smooth phase remains.}\label{rg_flow}
\end{figure*}

In this Letter, we investigate the strong coupling phases and the associated roughening transitions in a general class of noisy conserved growth models with an underlying chemical potential $\mu$ controlling the dynamics containing a nonlocal part. We demonstrate that when the nonlocal component of $\mu$ is weak, the surfaces can be both linearly and nonlinearly stable, or linearly unstable. However, at stronger nonlocality the surfaces [including CKPZ+ and MBE+ equations in two dimensions (2D)] can become nonlinearly unstable, even if it is linearly stable, signaling a transition to strong coupling phases that are argued to be crumpled for $d\leq d_c$, the critical dimension of the model. Our results, obtained by a combination of perturbative calculations, nonperturbative reasoning, and numerical studies, should provide a general theoretical framework to understand scaling, roughening and instabilities in  generic driven conserved growth processes including vapor deposition and idealized MBE.

We now proceed to derive these results, beginning with the equation of motion
  \begin{eqnarray}
&\hspace{-3.5 cm}
\frac{\partial h} {\partial t} =\bar\nu\nabla^2h-\nu\nabla^4 h - \frac{\lambda}{2}\nabla^2({\boldsymbol \nabla} h)^2\nonumber\\& \gap\quad
-\lambda_1\nabla_i\nabla_j(\nabla_i h \nabla_j h) + \eta. \label{mbe-ckpz}
\end{eqnarray}
In analogy with active membranes, we set $\nu > 0$  (analog of a bend modulus~\cite{nelson-book}), but similar to active tensions in active membranes, let $\overline \nu$ to be of any sign~\cite{sriram-PRL,debayan-aniso-kpz}. Both $\lambda$, $\lambda_1$ are nonlinear coupling constants with arbitrary signs of the lowest order symmetry-permitted active terms. Noise $\eta$ is zero mean, Gaussian distributed with a variance
\begin{equation}
\langle \hat \eta({\bf k},t)\hat \eta({\bf k'},t') \rangle = 2D |{\bf k}|^{-y}\delta^d({\bf k+k'}) \delta(t-t') \label{noise}
\end{equation}
in the Fourier space, where $\bf k,k'$ are wavevectors.
Here, $y\ge -2$ determines the spatial correlation of the noise. Equation~(\ref{mbe-ckpz}) implies a current $\mathbf{J}$ and a chemical potential $\mu$~\cite{ckpz_plus,mbe_plus} related via ${\bf J} = -{\boldsymbol \nabla} \mu$, where
\begin{eqnarray}
&\hspace{-2.5 cm}{\bf J}=-\bar\nu{\boldsymbol \nabla}h+\nu {\boldsymbol \nabla} \nabla^2 h
+ \frac{\lambda}{2} {\boldsymbol \nabla} [(\nabla h)^2]\nonumber \\& \gap\quad + \lambda_1 \nabla^{-2} {\boldsymbol \nabla}\left[\nabla_i\nabla_j(\nabla_i h \nabla_j h)\right].\label{current} 
\end{eqnarray}
The $\lambda_1$ term contributes nonlocally to $\mu$~\cite{ckpz,lds}.
If $\bar\nu\geq (<)0$, the model is linearly stable (unstable) in the long wavelength limit; $\bar\nu=0$ is akin to a (mean-field) critical point in equilibrium critical systems. One can define a ``correlation length'' $\bar\xi\equiv \sqrt{\nu/|\bar\nu|}$, such that for a surface of linear size $L<(>)\bar\xi$, the long wavelength properties are controlled by $\nu (\bar \nu)$. Focusing on the linearly stable case,  we note that the correlation function $C({\bf k},\omega)\equiv \langle |h({\bf k},\omega|^2\rangle= 2Dk^{-y}/[\omega^2 + (\bar\nu k^2 + \nu k^4)^2]$ in the linear theory. It now behooves us to find whether nonlinear effects are
relevant (in the RG or renormalization group sense). We apply the perturbative Wilsonian dynamic RG approach at one-loop order~\cite{stanley,forster,halpin}. Due to the form of (\ref{mbe-ckpz}), fluctuation corrections to the noise is at least ${\cal O}(k^4)$, whereas the noise variance is $\sim k^{-y},\,y\ge -2$, giving no relevant corrections to $D$; see, e.g., Refs.~\cite{yakhot,medina}. This leads to an exact exponent identity that holds in all order of perturbation theory (see below). At the largest scales, where relaxation is dominated by the $\bar\nu>0$-term in (\ref{mbe-ckpz}), there are no relevant fluctuation corrections { to $\bar\nu$ also}, giving the above result for $C$ exact in the asymptotic long wavelength limit. For sufficiently small $\bar\nu$, there should be a large enough range of scales in which the dominant relaxation is controlled by the $\nu$ term. Once again we apply RG,  which we remind the reader is only valid on length scales $L<\bar\xi$. This is implemented via a path integral formalism in terms of $h(\mathbf{x}, t)$ and its associated conjugate response field $\hat{h}(\mathbf{x}, t)$~\cite{supple} that is formally equivalent to Eq.~(\ref{mbe-ckpz}) together with the noise correlations in Eq.~(\ref{noise}).  Corrections to the parameters due to fluctuations are computed through one-loop Feynman diagrams (see the Supplemental Material (SM)~\cite{supple}). 
At the one-loop order, the nonlinear couplings $\lambda$ and $\lambda_1$ do not receive any relevant corrections. In contrast, $\nu$ acquires divergent corrections at one loop order. Dimensional analysis { gives} an effective dimensionless coupling constant $g$ having  $d_c=4+y$ and a dimensionless ratio $\gamma$ defined as
\begin{equation}
 g=k_d\frac{\lambda^2D}{\nu^3}\Lambda^{d_c-d},\;\gamma = \frac{\lambda_1}{\lambda},
\end{equation}
{ where $\Lambda$ is a wavevector scale.}
Using $d=4+y-\epsilon$ near $d_c$, the RG differential recursion relations ($\exp(l)$ is a dimensionless length) up to ${\cal O}(\epsilon)$ are
\begin{align}
   &\frac{dD}{dl}=D[z+y-d-2\chi],\label{D-flow} \\
   &\frac{d(\lambda,\lambda_1)}{dl}=(\lambda,\lambda_1)[\chi+z-4]\label{lam-flow},\\
   &\frac{d\nu}{dl}=\nu[z-4+g\Delta(\gamma,y)]\label{nu-flow},\\
   &\frac{dg}{dl}=g[\epsilon-3g\Delta(\gamma,y)],\label{g-flow}
   \end{align}
together with $d\gamma/dl=d/dl(\lambda_1/\lambda)=0$. Here,
\begin{align}
\Delta\equiv\frac{1}{2(4+y)}\biggl[1+\gamma(10+2y)+\gamma^2\biggl(\frac{8y+36}{6+y}\biggl)\biggl].\label{D_elta}
\end{align}
Furthermore, $z$ and $\chi$ are the dynamic and roughness exponents, respectively~\cite{stanley}; see also SM~\cite{supple}. 
{RG recursion relation (\ref{D-flow}) is exact, which implies $z+y-d-2\chi=0$ as an {\em exact} exponent identity in the present problem at all $d$; see also Refs.~\cite{lds,yakhot,medina}.}

At $d=d_c$, $\Delta(\gamma,y)>0$ (violet region in Fig.~\ref{rg_flow}(a)), $g=0$ is the only fixed point (which is stable) of (\ref{g-flow}), but its approach to 0 is so slow that $\nu$ still gets {\em infinitely renormalized}, giving logarithmic modulations of the  linear theory results $z=4$ and $\chi=0$. This is reminiscent of anomalous elasticity  in three-dimensional equilibrium smectics~\cite{smec1,smec2},
 2D equilibrium elastic sheet
having vanishing thermal expansion coupled with Ising spins~\cite{ising-el1,ising-el2} and generelized quasi-long-range order in an active, diffusive XY model~\cite{astik-xy1,astik-xy2}.
Width $\mathcal{W} \equiv \big\langle \big[h(\mathbf{x},t)-\overline{h}(t)\big]^2\big\rangle^{1/2}$ grows initially with time and eventually saturates to a steady value $\mathcal{W}_\mathrm{sat} \sim\left[\ln\left(L/a_0\right)\right]^{1/3}$. Here $\overline h(t)$ is the mean height at time $t$, $L$ is the system size and $a_0$ a microscopic cutoff. In spite of the one-loop perturbative origin of this result, it is in fact exact in the asymptotic long wavelength limit. In the present
theory, at the one-loop order $g(l ) \sim 1/l$ for large $l$ in the stable, weak coupling phase. Higher-loop corrections should entail corrections to this $1/l$ behavior, which vanish faster than $1/l$, giving only finite corrections to $\nu$, which are neglected in the spirit of RG. This together with the nonrenormalization of $D$ leaves the scaling of $\mathcal{W}_\mathrm{sat}$ with $L$ unchanged, a result valid to all orders in perturbation theory; see Refs.~\cite{astik-xy1,astik-xy2,debayan1} for similar arguments.
The white region in Fig.~\ref{rg_flow}(a) where $\Delta(\gamma,y)<0$ is the strong-coupling regime speculated to be unstable.
\begin{figure}[t]
 \includegraphics[width=\columnwidth]{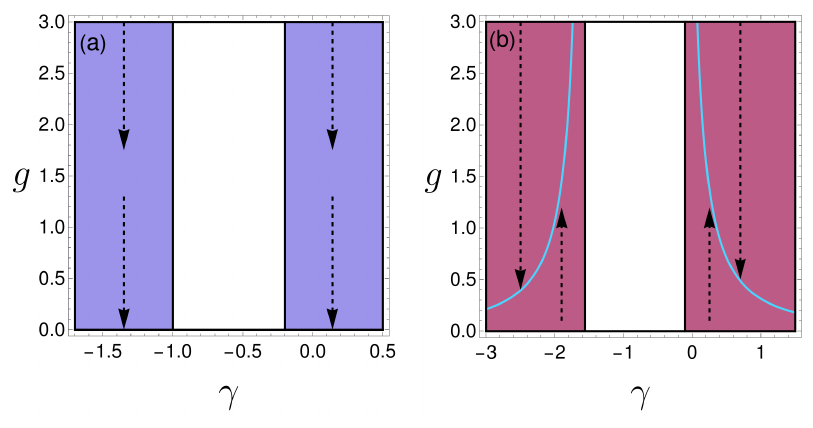}
 \caption{($\bar\nu=0$) RG flow diagram in 2D for the (a) CKPZ+ model. The violet region is the perturbatively accessible logarithmic rough phase, while the central white region is unstable. In the accessible region, the RG flow lines (indicated by arrows) flow toward the stable fixed line $g = 0$.~(b) MBE+ model. The blue line within the stable (red) region denotes the stable fixed line, toward which the RG flow lines converge.}\label{rg_flow_2d}
\end{figure}
The above results hold so long as the ``RG time'' $\ell^*$ at which the nonlinearities become important is less than $\ln (\bar\xi/a_0)$, such that any substantial renormalization of $\nu$ can already take place. An estimation of $\ell^*$ can be made by setting $\nu\sim\Delta\nu (l)$, the one-loop correction to $\nu$, giving  $\ell^*\sim 1/[g_0\Delta(\gamma,y)]$, where $g_0$ is the ``RG initial'' value of $g$.


To study the unstable regime more, we first consider the bare perturbation theory for $\nu$, which is the same as the RG calculations for $\nu$ except that now we extend the integrals over wave vectors down to an infrared cutoff $q_\text{min}=2\pi/L$  with $L<\bar\xi$. Setting $d=d_c$, we get 
\begin{eqnarray}
\nu_\text{eff} \approx \nu_0 +g_0\nu_0\Delta(\gamma,y)\ln (L/a_0), \label{nu-e}
\end{eqnarray}
for the effective $\nu$.
Thus, for $\Delta(\gamma, y) < 0$, we have $\nu_\text{eff} < \nu_0$ for sufficiently large system sizes. From (\ref{nu-e}) we get a length scale { $\xi\sim a_0\exp [1/g_0|\Delta(\gamma,y)|]$}, which is the analog of a de Gennes-Taupin persistence length for an equilibrium fluid membrane~\cite{taupin}, such that the variance of the local normal fluctuations $\langle ({\delta {\bf n}}^2)\rangle$ diverges for $L>\xi$, where {${\bf n}\approx (-\boldsymbol \nabla h, 1)\equiv (-\delta {\bf n}, 1)$} assuming  small fluctuations. {Thus orientational order, characterized by a finite $\langle ({\delta {\bf n}}^2)\rangle$, is lost as $L\rightarrow
\xi_-$, and the surface has only short range orientational order, a telltale signature of a crumpled surface.} Calculation of the effective membrane area $S=S_0\sqrt{1+({\boldsymbol\nabla} h)^2}$ relative to the base area $S_0$ gives, with $S\propto S_0^{\tilde d/2}$, $\tilde d/2\equiv d\ln S/d\ln S_0\simeq 1+ \langle ({\boldsymbol\nabla} h)^2\rangle/2$ grows sharply as $L\rightarrow \xi_-$, indicating that not only the surface gets more flexible due to the reduction in $\nu$, it also gets highly corrugated for  $L\rightarrow \xi_-$.  {Indeed, since $\langle (\boldsymbol\nabla h)^2\rangle\sim (D/\nu_\text{eff})\times {\cal O}(1)$, which diverges as $\nu_\text{eff}$ vanishes for $L\rightarrow \xi_-$, $\tilde d$ also diverges in the same limit.} This strongly suggests that for parameters in the white region in Fig.~\ref{rg_flow}(a), the surface should appear crumpled; see Refs.~\cite{peliti,tirtha-mem-pre} for similar studies on crumpling of fluid membranes.

We now focus on { the weak-coupling ``ordered'' phase} at $d<d_c$. Using the RG flow equations for $d<d_c$ ($\epsilon>0$), when $\Delta(\gamma,y)>0$, the plane
$g^*=\epsilon/\left(3\Delta(\gamma,y)\right)$ [blue plane in Fig.~\ref{rg_flow}(b)] is a stable fixed plane in
the $\gamma$-$y$-$g$ space, giving $z=4-\epsilon/3$ and $\chi=\epsilon/3>0$ to ${\cal O}(\epsilon)$,
while $g^*=0$ is unstable. For $\Delta(\gamma,y)<0$ [white region in Fig.~\ref{rg_flow}(b)], $g^*=0$ is the only fixed
plane but unstable, indicating a strong-coupling (crumpled) phase
\cite{mbe_plus}. For $d>d_c$ ($\epsilon<0$) and $\Delta(\gamma,y)<0$,
$g^*=0$ is a stable fixed plane spanned by $\gamma$-$y$-$g$ with linear theory exponents $z=4$,
$\chi=-|\epsilon|/2<0$, bare perturbative corrections to $\nu_0$ are finite, while
$g^*=|\epsilon|/\left(3|\Delta(\gamma,y)|\right)$ [blue plane in Fig.~\ref{rg_flow}(c)] is unstable, signaling a
roughening transition akin to the KPZ equation~\cite{stanley,mbe_plus}. For $\Delta(\gamma,y)>0$ and
$\epsilon<0$, $g^*=0$ remains the only stable plane, giving a smooth phase [green region in Fig.~\ref{rg_flow}(c)]. With $\epsilon<0$, if $\Delta(\gamma,y)>(<)0$, $g(l)$ approaches (moves away from) the stable (unstable) fixed point $g^*=0\ \left(|\epsilon|/(3|\Delta(\gamma,y)|)\right)$ of (\ref{g-flow}) {\em exponentially fast} [diverges at a finite RG scale $l_c$ as $g(l_c)\sim\mathcal{O}(1)$, beyond which perturbative RG breaks down], thus identifying $d_c=4+y$ as the {\em upper} ({\em lower}) critical dimension; see also Refs.~\cite{debayan1,astik-xy1,astik-xy2}. {While the individual values of $z$ and $\chi$ are obtained only at ${\cal O}(\epsilon)$ for $d<d_c$, the exponent identity from (\ref{D-flow}) continue to hold. } 

{The roughening transition at $d<d_c$ between the weak-coupling and strong-coupling unstable phases here thus signifies the loss of long-range orientational order, with the surface being statistically flat only in the weak coupling phase. This distinguishes it from its analog in the KPZ equation for $d>2$, where long-range positional order {that is characterized by a finite variance of the local height fluctuations in the large system size limit,} is lost at the roughening transition. Thus the roughening transition here is associated with a loss of symmetry {\em different} from its KPZ analog, which is detectable in experimental or numerical measurements of surface conformations.}

We briefly revisit the $d=2$ cases: CKPZ+ ($y=-2$) and MBE+ ($y=0$), studied in Refs.~\cite{ckpz_plus,mbe_plus}. For CKPZ+, with $d_c=2$, bare perturbation theory predicts instability for $-1.0<\gamma<-0.2$ [white region in Fig.~\ref{rg_flow_2d}(a)], while the stable phase (violet) shows linear exponents $z=4$, $\chi=0$, and logarithmic roughening. For MBE+ [Fig.~\ref{rg_flow_2d}(b)], the red region is stable with a fixed line $g^*=2/\!\left(3\Delta(\gamma)\right)$, with $\Delta(\gamma)=\tfrac{3}{4}\gamma^2+\tfrac{5}{4}\gamma+\tfrac{1}{8}$, giving $z=10/3$, $\chi=2/3$, consistent with the LDS class~\cite{lds}. The white region ($-1.559<\gamma<-0.106$) is believed to be unstable or crumpled~\cite{mbe_plus}.

This instability can be cut off by the $\bar\nu \nabla^2 h$ in (\ref{mbe-ckpz}) in the long wavelength limit, if the ``RG time'' $l_c$ at which $g(l_c)$ becomes ${\cal O}(1)$ is substantially bigger than $\ln(\xi/a_0)$. Setting $g(l_c)\sim 1$ gives another ``correlation length'' $\xi\sim a_0\exp (l_c)\sim \exp[(1-g_0)/3g_0|\Delta(\gamma,y)|]$. If $\bar\xi<\xi$, the long wavelength behavior is controlled by $\bar\nu$ and $D$, which due to  the absence of any fluctuation-corrections for both $\bar\nu, D$ give $z=2$ and $\chi=(2+y-d)/2$, otherwise for $\bar\xi>\xi$, the surface gets crumpled before the stabilizing $\bar\nu$ term becomes important.

To further analyze the unstable regions observed in the RG flow diagrams, we now employ a one-loop MCT calculation~\cite{jkb-mct,bmhd2,akc-ab-jkb,abmhd,abjkb} to extract the associated scaling behavior for $\bar\nu\approx0$. In the present problem, due to the absence of any diverging correction to the noise amplitude (an exact statement) and the lack of relevant vertex corrections at the one-loop order,  MCT entails only solving $G$ self-consistently. The response function $G({\bf k}, \omega)$ has the scaling form
\begin{align}
&G({\bf k},\omega)=k^{-z}g\biggl(\frac{\omega}{k^z}\biggr)=\frac{1}{-i\omega+\Sigma({\bf k},\omega)},\label{resp1}
\end{align}
where the self-energy $\Sigma ({\bf k},\omega=0)=\Gamma k^z$ in the zero frequency limit. The correlator is constructed in the Lorentzian  approximation valid for low $\omega$~\cite{supple,debayan2}.

By combining the contributions from all the relevant one-loop diagrams for $G^{-1}({\bf k}, \omega = 0)$ and assuming the dominant contribution comes from the infrared limit, we arrive at the dominant one-loop contribution to the self-energy~\cite{supple} $k_d\frac{\lambda^2 D}{\Gamma^2}\tilde\Delta_m(\gamma,y,z)k^4\frac{k^{4+d-y-2z}}{y+2z-4-d}$, where
\begin{align}
&\tilde\Delta_m (\gamma,y,z)=\biggl[\frac{2+y+z-d}{4d}+\gamma^2\biggl(\frac{1}{d}+\frac{3(y+z)}{d(d+2)}\biggl)\nonumber\\ & \gap+\gamma\biggl(\frac{y+z+d-1}{2d}+\frac{3(2+y+z)}{2d(d+2)}\biggl)\biggl],\label{mct_eq1}
\end{align}
assuming $\tilde\Delta_m(\gamma,y,z)>0$ and $4+d-y-2z<0$.
\begin{figure}[t]
\centering
\includegraphics[width=0.49\textwidth]{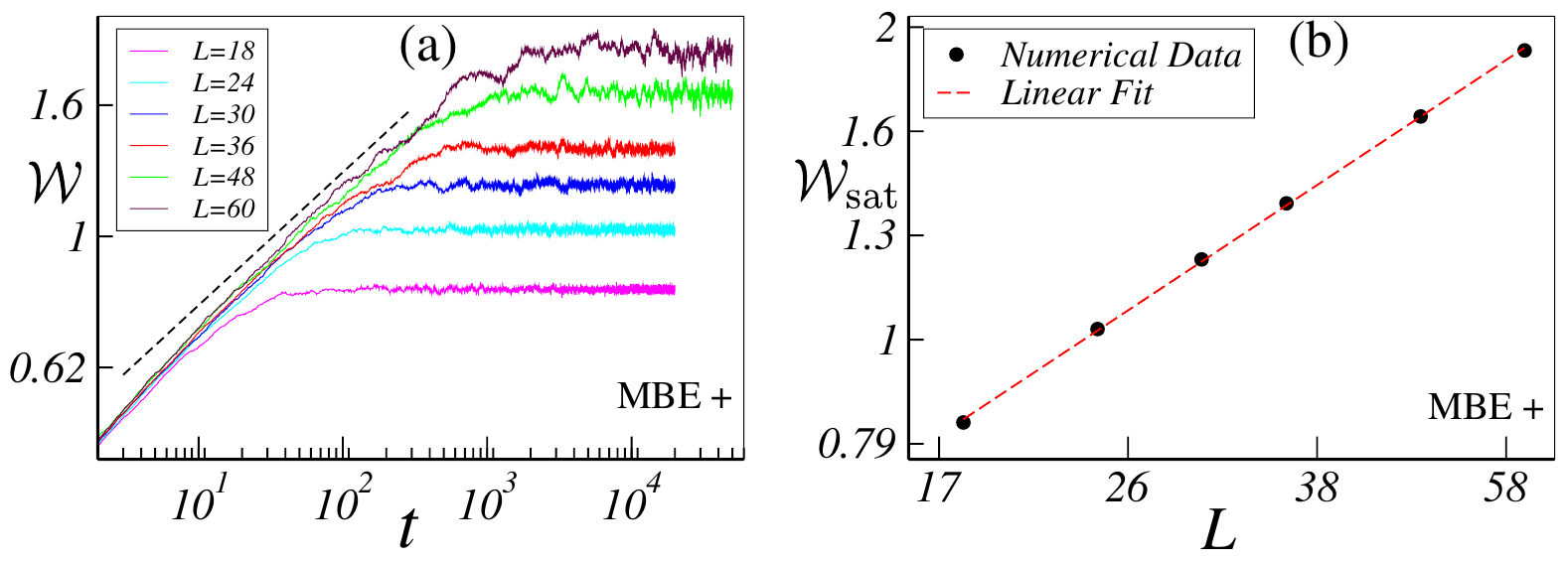}
\caption{For the MBE+ model in $d=2$, log-log plots [with $ \bar\nu=0, \nu=1$, $D=1$, $\nu_r=0.2$, $\lambda=2$, $\lambda_1 = 1$, and time step $(\delta t) = 10^{-3}$] of (a) ${\mathcal W}$ versus $t$. The black dashed line indicates a linear fit in the growth regime, yielding $\beta = 0.207 \pm 7.26 \times 10^{-5}$. (b) ${\mathcal W}_{\text{sat}}$ versus $L$. The red dashed line represent linear fit to the numerical data displayed on a log-log scale, with slope $\chi = 0.707 \pm 0.006$. { These numerical results were obtained using the Euler scheme for time integration.}  }
\label{plot_mbe_stable}
\end{figure}
With $\epsilon>0$ ($d<d_c$) and $\tilde\Delta_m>0$, matching $k$ powers with the bare term $\Gamma k^z$ yields $z=4-\epsilon/3$ and $\chi=\epsilon/3$ (using $\chi+z=4$), consistent with RG. For $\tilde\Delta_m<0$, MCT fails, indicating a crumpled phase. This is distinct from the MCT of KPZ equation, see, e.g.,~\cite{healy_frg,jkb-mct,jkb-book,bmhd1,bmhd2}, in which MCT {does not} fail in the strong coupling phase. Thus, the strong coupling phases in (\ref{mbe-ckpz}) is clearly distinct from that in the KPZ equation. At $d=d_c$ we get $z=4$, with one-loop correction to the self-energy~\cite{supple},
$k_d\frac{\lambda^2 D}{\Gamma^2}\ln(\Lambda/k)\Delta_m(\gamma,y)$, with
\begin{align}
\Delta_m = \frac{1}{2(4+y)}\biggl[1+\gamma(10+2y)+\gamma^2\biggl(\frac{8y+36}{6+y}\biggl)\biggl].
\end{align}
\begin{figure}[b]
\centering
\includegraphics[width=0.49\textwidth]{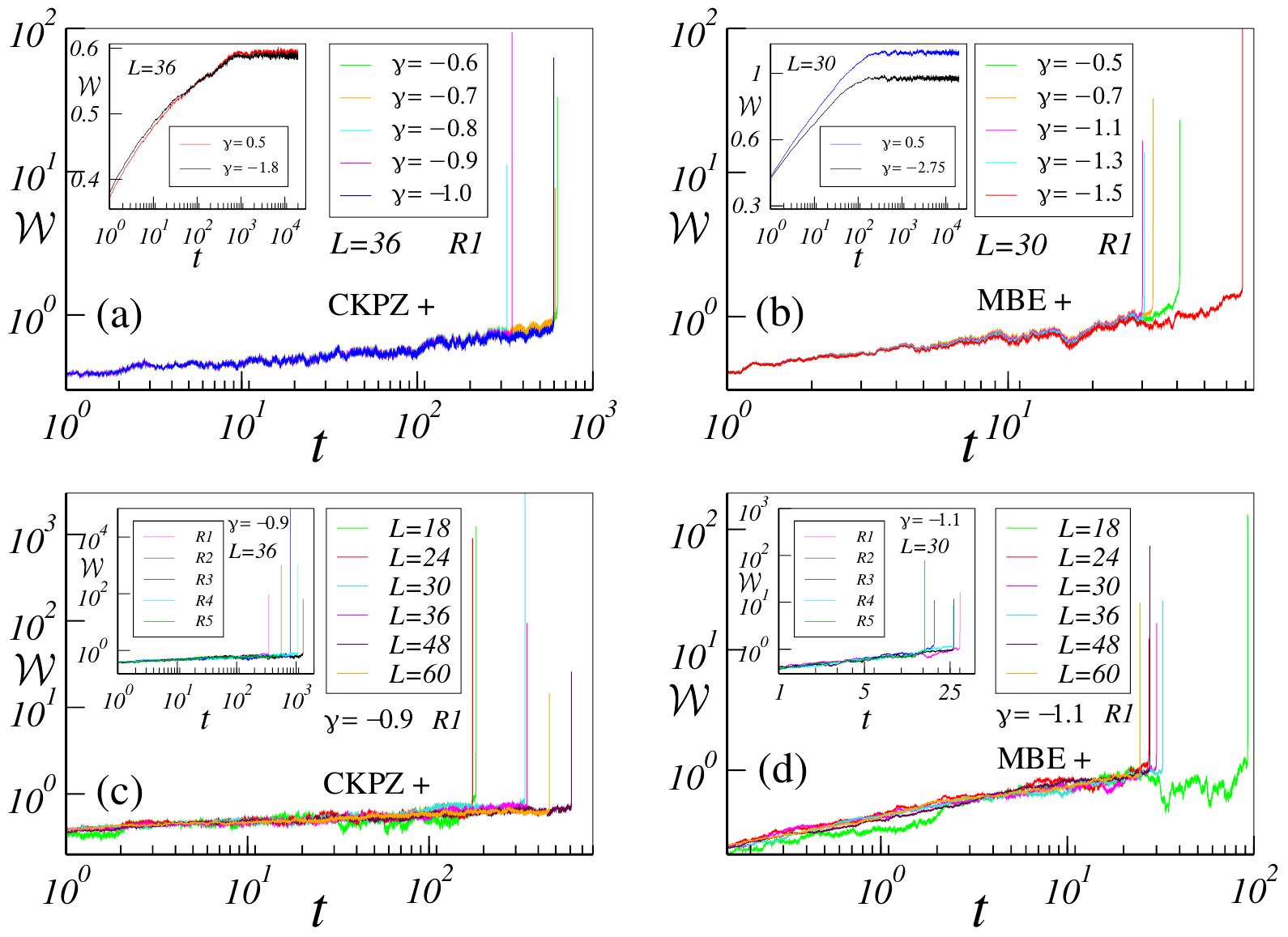}
\caption{Log-log plots of ${\mathcal W}$ versus $t$ for various values of $\gamma$ in the unstable region of Fig.~\ref{rg_flow_2d}, for a fixed realization ($R1$) and fixed $L$. Inset: log-log plots of ${\mathcal W}$ versus $t$ for various values of $\gamma$ in the stable region of Fig.~\ref{rg_flow_2d}, where ${\mathcal W}$ (averaged over many realizations) saturates, for the (a) CKPZ+ model and (b) MBE+ model. Log-log plots of ${\mathcal W}$ versus $t$ for various system sizes but fixed realization and fixed $\gamma$ in the unstable region. Inset: same but for a fixed $L$, fixed $\gamma$, and over different realizations, for the (c) CKPZ+ model and (d) MBE+ model. Parameters $\bar\nu=0$, $\nu=1$, $D=1$, $\nu_r=0.2$, $\lambda=2$, with various values of $\lambda_1$ (i.e., $\gamma$), and time step $\delta t = 10^{-3}\,(10^{-4})$ in the stable (unstable) region. { These numerical results were obtained using the Euler scheme for time integration.}}
\label{plot_unstable}
\end{figure}
The obtained $\Delta_m(\gamma, y)$  exactly matches the RG result $\Delta(\gamma, y)$ in Eq.~(\ref{D_elta}). The MCT approach breaks down for $\Delta(\gamma, y) < 0$, signaling an instability or crumpling of the surface; see the white region in Fig.~\ref{rg_flow}(a). This suggests that the model is no longer valid in this region. 
With $\Delta(\gamma, y) > 0$, the model remains stable, yielding the scaling exponents $z = 4$ and $\chi = 0$ in agreement with RG.
For the CKPZ+ model ($y=-2$) $d=d_c=2$, MCT predicts the model is unstable in the range $-1.0 < \gamma < -0.2$, consistent with RG results [Fig.~\ref{rg_flow_2d}(a)]. For the MBE+ model ($y=0$), MCT identifies an unstable window $-0.936 < \gamma < -0.254$. Outside this range, $z=10/3$ and $\chi=2/3$, in agreement with RG [Fig.~\ref{rg_flow_2d}(b)]. The slight mismatch between MCT and RG unstable regions arises from the $\epsilon$ expansion used in RG. Independent of whether the surface remains stable or not for length scales $<\xi$, with $\bar\nu<0$ there is long wavelength instability that would dominate for a sufficiently large $L>\xi$.

To validate our results, we perform direct numerical solutions (DNS) for both CKPZ+ and MBE+ models at $d=2$ with $\bar\nu=0$. We include a regularizing term $\nu_r \nabla^6 h,\,\nu_r>0$ in Eq.~(\ref{mbe-ckpz}) to control numerical instabilities~\cite{ckpz_plus,abmhd_prl,abssr,sahoo}. { We have used the pseudospectral method with the $2/3$ dealiasing rule and Euler~\cite{euler} scheme for time integration. }Figure~\ref{plot_mbe_stable}(a) shows $\mathcal{W}$ vs $t$ for the MBE+ model at $\gamma=0.5$ [stable region of Fig.~\ref{rg_flow_2d}(b)] for different system sizes, averaged over several realizations, yielding $\beta = 0.207 \pm 7.26 \times 10^{-5}$. Figure~\ref{plot_mbe_stable}(b) displays the variation of $\mathcal{W}_\text{sat}$ with $L$, respectively, giving  $\chi = 0.707 \pm 0.006$. These results confirm algebraic scaling in the MBE+ model, in close agreement with RG predictions $\chi = 2/3$. Numerical results for CKPZ+ are known~\cite{ckpz_plus}, with stable-region details given in the SM~\cite{supple}. To investigate the unstable regime, we fix $\delta t=10^{-4}$, set $\lambda=2$, and vary $\lambda_1$ to tune $\gamma$. The term $\nu_r \nabla^6 h$ stabilizes short-wavelength modes but is irrelevant in the unstable region $\nu<0$. For $L > L_c \sim \sqrt{\nu_r/|\nu|} \approx 0.45$, height and normal fluctuations diverge, indicating possible membrane crumpling~\cite{sm}. Figures~\ref{plot_unstable}(a) and~\ref{plot_unstable}(b) show $\mathcal{W}(t)$ for CKPZ+ (MBE+) models with {$L=36\,(30)$} across several $\gamma$ values in the unstable region of Fig.~\ref{rg_flow_2d}, where $\mathcal{W}$ diverges without saturation. Insets show $\gamma$ values from the stable domains, where $\mathcal{W}$ saturates, confirming the white region as unstable. Figures~\ref{plot_unstable}(c) and~\ref{plot_unstable}(d) show $\mathcal{W}(t)$ for varying $L$ at fixed $\gamma$ in the unstable regime, with insets displaying different realizations for fixed $L$, $\gamma$ for CKPZ+ (MBE+) models. The consistently diverging $\mathcal{W}$ confirms our speculations that this region is unstable. This agrees with previous studies of the deterministic CKPZ+ equation in 2D, which reported blowup solutions for particular combinations of $\lambda$ and $\lambda_{1}$~\cite{unstable_1,unstable_2}. The robustness of the runaway behavior giving uncontrolled fluctuations in the putative crumpled phase is further confirmed from additional DNS studies using a higher-order Heun scheme~\cite{euler} for time integration and varying $\nu_r$, which are in agreement with the Euler scheme results; see the SM~\cite{supple}.


To summarize, we have studied a generalized conserved surface growth model that in two distinct limits reduces to the CKPZ+ and MBE+ equations. In the strong coupling phases of this equation, achieved for sufficiently strong nonlocal contributions to the chemical potential, the surface is unstable. We use a variety of techniques, analytical and numerical, to argue that this unstable surface is crumpled. This makes the strong coupling phases in this model and the associated roughening transition distinct from those in the KPZ equations in $d>2$. Nonrenormalization of $\lambda$ in the present model holds only at the one-loop order~\cite{janssen}. While this can potentially make our results on the scaling exponents quantitatively inaccurate, our predictions on crumpling, corroborated by DNS studies, should generally hold. {While we predict a crumpling transition to a possible crumpled phase,  statistical properties in the crumpled phase cannot be investigated directly from (\ref{mbe-ckpz}), since large fluctuations in a putative crumpled phase are likely to make use of the Monge gauge, in which (\ref{mbe-ckpz}) is constructed, questionable. }
It will be interesting to study the spatiotemporal fluctuations of a conserved density on a conserved fluctuating surface as here~\cite{nilu1,nilu2,tirtha-jstat,sm1,sm2} and the consequence of coupling of a conserved surface with an elastic network~\cite{tethered1,tethered2}, or with quenched disorder~\cite{sm-ckpz}. Construction of appropriate lattice models that include microscopic dynamical processes corresponding to nonlocal chemical potentials by extending those studied in Refs.~\cite{latt1,latt2,latt3} should be a promising research direction. While we are not aware of any direct experimental realizations with long range noises yet, ion sputtering phenomena~\cite{ion} with spatially long-range correlated ion bombardment could be a promising possibility. {Experimental detection of overhangs, which can affect the quality or usefulness of a film produced, should indicate crumpling, which may be resolved in direct imaging; see, e.g., Ref.~\cite{imaging}, although a direction of overhangs with crumpling requires additional work beyond the scope of this study. We expect our results will stimulate further research in this direction.}

{\em Acknowledgments.} The authors thank A.~Haldar, T. Halpin-Healy and M. Kardar for comments and helpful suggestions. A.B. thanks the AvH Stiftung (Germany) for partial financial support under the Research Group Linkage Programme scheme (2024). A.B. thanks ANRF (India) for partial financial support through the ARG (MATRICS) programme (File No. ANRF/ARGM/2025/000461/TS).

\bibliography{ckpz_mbe.bib}

\begin{thebibliography}{78}
\bibitem [{sup()}]{supple}%
  \BibitemOpen
  \bibfield  {title} {\enquote {\bibinfo {title} {\uppercase{S}ee
  \uppercase{S}upplemental \uppercase{M}aterial for the one-loop
  \uppercase{F}eynman diagrams used in the renormalization group analysis,
  details of the \uppercase{MCT} calculations, and numerical method, which
  includes \uppercase{R}efs.~\cite{bausch, tauber, yakhot, ckpz_plus,
  euler}},}\ }\href@noop {} {\ }\BibitemShut {NoStop}%
\end{thebibliography}

\clearpage
\onecolumngrid
\begin{center}
\textbf{\LARGE Supplemental Material}
\end{center}
\vspace{0.5cm}

\section{Renormalization group calculations}


We start by constructing a generating functional~\cite{bausch,tauber} by using Eq.~(1) and Eq.~(2) of the main text. We get
\begin{equation}
 \mathcal{Z}=\int \mathcal{D}\hat{h} \mathcal{D}h e^{-\mathcal{S}[\hat{h},h]},
\end{equation}
where $\hat{h}$ is referred as response field and $\mathcal{S}$ is action functional, which is given by
\begin{align}
S= -\int_{{\bf k},t}|{\bf k}|^{-y}D|\hat{h}|^2 + \int_{x,t}\hat{h}(\partial_th-\bar\nu\nabla^2 h+ \nu\nabla^4h +\frac{\lambda}{2} \nabla^2(\boldsymbol\nabla h)^2+\lambda_1\nabla_i\nabla_j(\nabla_ih \nabla_j h)).\label{action}
\end{align}


\subsection{Results from the harmonic theory or the linearized equation of motion}

This can be obtained from the Gaussian part of the action functional~(\ref{action}), i.e., by setting $\lambda=\lambda_1=0$ in (\ref{action}).
\begin{figure}[b]
\centering
\includegraphics[width=0.30\textwidth]{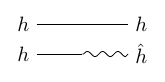}
\caption{Diagrammatic representations of two point functions.}
\label{propagator}
\end{figure}
We first define Fourier transforms in space and time by
\begin{equation*}
h({\bf x},t)=\int_{{\bf k},\omega}h({\bf k},\omega)e^{i({\bf k\cdot x}-\omega t)}.
\end{equation*}
Then the (bare) two-point functions in the Gaussian limit of (\ref{action}) are
\begin{subequations}
\begin{align}
&\langle \hat{h}({\bf k},\omega) \hat{h}(-{\bf k},-\omega)\rangle_0=0,\\
&\langle \hat{h}({\bf k},\omega) h(-{\bf k},-\omega)\rangle_0=\frac{1}{i\omega +\bar\nu k^2 +\nu k^4},\\
&\langle \hat{h}(-{\bf k},-\omega) h({\bf k},\omega)\rangle_0=\frac{1}{-i\omega +\bar\nu k^2+\nu k^4},\\
&\langle h({\bf k},\omega) h(-{\bf k},-\omega)\rangle_0=\frac{2D|{\bf k}|^{-y}}{\omega^2 +(\bar\nu k^2+\nu k^4)^2}.
\end{align}
\end{subequations}
Fig.~\ref{propagator} shows the diagrammatic representations of the various  two-point functions.



\subsection{One-loop corrections to $\nu$}

\begin{figure}[htb]
\includegraphics[width=0.44\textwidth]{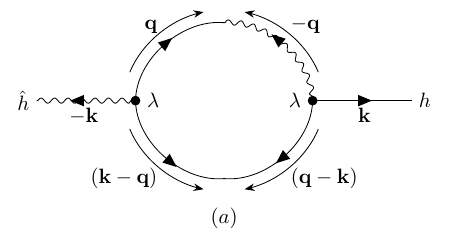}
\includegraphics[width=0.44\textwidth]{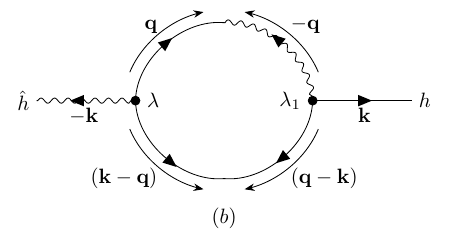}
\includegraphics[width=0.44\textwidth]{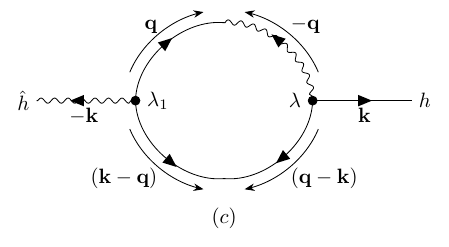}
\includegraphics[width=0.44\textwidth]{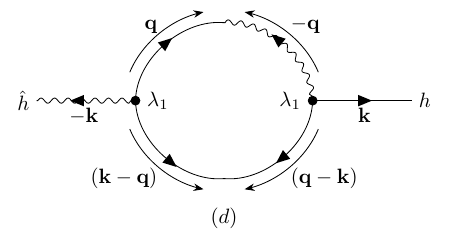}
\caption{ One-loop Feynman diagrams that contribute to the renormalization of $\nu$}
\label{nu_diag}
\end{figure}

We provide the details of the one-loop RG calculations for $\bar\nu=0$.
The one-loop Feynman diagrams, that contribute to the fluctuation correction of $\nu$ is shown in Fig.~\ref{nu_diag}, for various vertex combinations.
Figure~\ref{nu_diag}(a) has a symmetry factor of 8. Its contribution is given by
\begin{align}
\lambda^2\int_{{\bf q},\Omega}k^2q_i(k-q)_iq^2(q-k)_jk_j\times\frac{1}{-i\Omega+\nu q^4}\times\frac{2D|{\bf k-q}|^{-y}}{\Omega^2+\nu^2(k-q)^8}.
\end{align}
After performing the $\Omega$-integral and substituting
$$\int_{-\infty}^{\infty}\frac{d\Omega}{2\pi}\frac{1}{(-i\Omega+\nu q^4)(\Omega^2+\nu^2(k-q)^8)}=\frac{1}{2\nu^2(k-q)^4[(k-q)^4+q^4]}$$
in above equation, we obtain,
\begin{align}
\frac{\lambda^2 D}{\nu^2}(k^2k_j)\int_{{\bf q}}\frac{q_i(k-q)_iq^2(q-k)_j}{|{\bf k-q}|^{4+y}[(k-q)^4+q^4]}.\label{aa}
\end{align}
After symmetrizing, i.e., making ${\bf q}\rightarrow {{\bf q}+{{\bf k}}/{2}}$, the numerator and denominator separately become
\begin{align*}
&q_i(k-q)_iq^2(q-k)_j=-q^4q_j+q^4\frac{k_j}{2}-q^2q_jq_mk_m,\\
&\frac{1}{|{\bf k-q}|^{4+y}[(k-q)^4+q^4]}=\frac{1}{2q^{8+y}}\Bigl[1+\Bigl(\frac{4+y}{2}\Bigl)\frac{{\bf k \cdot q}}{q^2}\Bigl].
\end{align*}
Substituting these in Eq.~(\ref{aa}) we obtain three integrals
\begin{align}
&\mathcal {I}_1=\frac{\lambda^2 D}{\nu^2}(k^2k_j)\int_{\bf q}\frac{-q^4q_j}{2q^{8+y}}\Bigl(\frac{4+y}{2}\Bigl)\frac{k_mq_m}{q^2}=-\frac{\lambda^2 D}{\nu^2}\Bigl(\frac{4+y}{4d}\Bigl)k^4\int\frac{d^dq}{q^{4+y}},\\
&\mathcal {I}_2=\frac{\lambda^2 D}{\nu^2}(k^2k_j)\int_{\bf q}\frac{q^4k_j}{4q^{8+y}}=\frac{\lambda^2 D}{\nu^2}\frac{1}{4}k^4\int\frac{d^dq}{q^{4+y}},\\
&\mathcal {I}_3=\frac{\lambda^2 D}{\nu^2}(k^2k_j)\int_{\bf q}\frac{-q^2q_jq_mk_m}{2q^{8+y}}=-\frac{\lambda^2 D}{\nu^2}\frac{1}{2d}k^4\int\frac{d^dq}{q^{4+y}}.
\end{align}
Here, $\mathcal {I}_1$, $\mathcal {I}_3$ is evaluated by using the well-known relation~\cite{yakhot}
\begin{align}
k_ik_j\int d^dqf(q^2)q_iq_j=k_ik_j\times\frac{[\delta_{ij}]}{d}\int d^dq f(q^2)q^2.\label{identity_1}
\end{align}
Adding above three integrals we obtain correction to $\nu$ from vertex ($\lambda$-$\lambda$) and it is
\begin{align}
-\frac{\lambda^2 D}{\nu^2}k^4k_d\Bigl(\frac{6+y-d}{4d}\Bigl)\int q^{d-y-5} dq.\label{nu_corr_1}
\end{align}
Here $k_d=\frac{S_d}{(2\pi)^d}$, where $S_d$ is the surface area of a $d$-dimensional unit sphere. Fig.~\ref{nu_diag}(b) has a symmetry factor of 4. Its contribution is given by
\begin{align}
&2\lambda\lambda_1\int_{{\bf q},\Omega}k^2{\bf q\cdot (k-q)}q_iq_jk_i(q-k)_j\times\frac{1}{-i\Omega+\nu q^4}\times\frac{2D|{\bf k-q}|^{-y}}{\Omega^2+\nu^2(k-q)^8},\\
&=\frac{2\lambda\lambda_1 D}{\nu^2}(k^2k_i)\int_{{\bf q}}\frac{q_m(k-q)_mq_iq_j(q-k)_j}{|{\bf k-q}|^{4+y}[(k-q)^4+q^4]}.
\end{align}
After symmetrization, i.e., making ${\bf q}\rightarrow {{\bf q}+{{\bf k}}/{2}}$, the above integral can be written as the sum of two separate integrals. These are
\begin{align}
&\mathcal {I}_4=\frac{2\lambda\lambda_1 D}{\nu^2}(k^2k_i)\int_{\bf q}\frac{-q^4q_i}{2q^{8+y}}\Bigl(\frac{4+y}{2}\Bigl)\frac{k_mq_m}{q^2}=-\frac{2\lambda\lambda_1 D}{\nu^2}\Bigl(\frac{4+y}{4d}\Bigl)k^4\int\frac{d^dq}{q^{4+y}},\label{i4}\\
&\mathcal {I}_5=\frac{2\lambda\lambda_1 D}{\nu^2}(k^2k_i)\int_{\bf q}\frac{-q^4k_i}{4q^{8+y}}=-\frac{2\lambda\lambda_1 D}{\nu^2}\frac{1}{4}k^4\int\frac{d^dq}{q^{4+y}}.\label{i5}
\end{align}
Similarly, Fig.~\ref{nu_diag}(c) has a symmetry factor of 4 and its contribution is
\begin{align}
&2\lambda\lambda_1\int_{{\bf q},\Omega}k_ik_jq_i(k-q)_jq^2{\bf k\cdot (q-k)}\times\frac{1}{-i\Omega+\nu q^4}\times\frac{2D|{\bf k-q}|^{-y}}{\Omega^2+\nu^2(k-q)^8},\\
&=\frac{2\lambda\lambda_1 D}{\nu^2}(k_ik_jk_m)\int_{{\bf q}}\frac{q_i(k-q)_jq^2(q-k)_m}{|{\bf k-q}|^{4+y}[(k-q)^4+q^4]}.
\end{align}
After symmetrization, i.e., making ${\bf q}\rightarrow {{\bf q}+{{\bf k}}/{2}}$, the above integral integral can be written as the sum of three separate integrals. These are
\begin{align}
&\mathcal {I}_6=\frac{2\lambda\lambda_1 D}{\nu^2}(k_ik_jk_m)\int_{\bf q}\frac{-q_iq_jq^2q_m}{2q^{8+y}}\Bigl(\frac{4+y}{2}\Bigl)\frac{k_nq_n}{q^2}=-\frac{2\lambda\lambda_1 D}{\nu^2}\frac{3}{d(d+2)}\Bigl(\frac{4+y}{4}\Bigl)k^4\int\frac{d^dq}{q^{4+y}},\label{i6}\\
&\mathcal {I}_7=\frac{2\lambda\lambda_1 D}{\nu^2}(k_ik_jk_m)\int_{\bf q}\frac{q_iq_jq^2k_m}{4q^{8+y}}=\frac{2\lambda\lambda_1 D}{\nu^2}\frac{1}{4d}k^4\int\frac{d^dq}{q^{4+y}},\label{i7}\\
&\mathcal {I}_8=\frac{2\lambda\lambda_1 D}{\nu^2}(k_ik_jk_m)\int_{\bf q}\frac{-q_iq_jq_mq_nk_n}{2q^{8+y}}=-\frac{2\lambda\lambda_1 D}{\nu^2}\frac{3}{2d(d+2)}k^4\int\frac{d^dq}{q^{4+y}}.\label{i8}
\end{align}
Here in evaluating $\mathcal {I}_6$, $\mathcal {I}_8$,  we have used the following identity~\cite{yakhot}
\begin{align}
&k_ik_jk_mk_n\int d^dqf(q^2)q_iq_jq_mq_n=k_ik_jk_mk_n\frac{[\delta_{ij}\delta_{mn}+\delta_{im}\delta_{jn}+\delta_{in}\delta_{jm}]}{d(d+2)}\int d^dq f(q^2)q^4.\label{identity_2}
\end{align}
Adding Eqs.~(\ref{i4}), (\ref{i5}) and (\ref{i6})-(\ref{i8}) we obtain correction to $\nu$ from vertex ($\lambda$-$\lambda_1$) and it is
\begin{align}
-\frac{\lambda\lambda_1 D}{\nu^2}k^4k_d\Bigl(\frac{(3+d+y)}{2d}+\frac{3(6+y)}{2d(d+2)}\Bigl)\int q^{d-y-5} dq.\label{nu_corr_2}
\end{align}

Lastly, Fig.~\ref{nu_diag}(d) has a symmetry factor of 8 and its contribution is
\begin{align}
&4\lambda_1^2\int_{{\bf q},\Omega}k_ik_jq_i(k-q)_jq_mq_nk_m(q-k)_n\times\frac{1}{-i\Omega+\nu q^4}\times\frac{2D|{\bf k-q}|^{-y}}{\Omega^2+\nu^2(k-q)^8},\\
&=\frac{4\lambda_1^2 D}{\nu^2}(k_ik_jk_m)\int_{{\bf q}}\frac{q_i(k-q)_jq_mq_n(q-k)_n}{|{\bf k-q}|^{4+y}[(k-q)^4+q^4]}.
\end{align}
After symmetrization, i.e., ${\bf q}\rightarrow {{\bf q}+{{\bf k}}/{2}}$, the above integral reduces to two separate integrals, which are
\begin{align}
&\mathcal {I}_9=\frac{4\lambda_1^2 D}{\nu^2}(k_ik_jk_m)\int_{\bf q}\frac{-q^2q_iq_jq_m}{2q^{8+y}}\Bigl(\frac{4+y}{2}\Bigl)\frac{k_nq_n}{q^2}=-\frac{4\lambda_1^2 D}{\nu^2}\frac{3}{d(d+2)}\Bigl(\frac{4+y}{4}\Bigl)k^4\int\frac{d^dq}{q^{4+y}},\label{i9}\\
&\mathcal {I}_{10}=\frac{4\lambda_1^2 D}{\nu^2}(k_ik_jk_m)\int_{\bf q}\frac{-q^2q_iq_jk_m}{4q^{8+y}}=-\frac{4\lambda_1^2 D}{\nu^2}\frac{1}{4d}k^4\int\frac{d^dq}{q^{4+y}}.\label{i10}
\end{align}
Adding Eqs.~(\ref{i9}), (\ref{i10}) we obtain correction to $\nu$ from vertex ($\lambda_1$-$\lambda_1$) and it is
\begin{align}
-\frac{\lambda_1^2 D}{\nu^2}k^4k_d\Bigl(\frac{3(4+y)}{d(d+2)}+\frac{1}{d}\Bigl)\int q^{d-y-5} dq.\label{nu_corr_3}
\end{align}
Collecting all the corrections to $\nu$ i.e., adding Eqs.~(\ref{nu_corr_1}), (\ref{nu_corr_2}) and (\ref{nu_corr_3}) we obtain total correction to $\nu$
\begin{align}
\nu^<=\nu\Biggl[1+\frac{\lambda^2 D}{\nu^3}k_d\biggl[\frac{6+y-d}{4d}+\gamma\biggl(\frac{3+d+y}{2d}+\frac{3(6+y)}{2d(d+2)}\biggl)+\gamma^2\biggl(\frac{3(4+y)}{d(d+2)}+\frac{1}{d}\biggl)\biggl]\int_{\Lambda/b}^{\Lambda} q^{d-y-5} dq\Biggl].\label{nu_corr_total}
\end{align}
Here, $\gamma=\frac{\lambda_1}{\lambda}$, as defined in the main text.

\subsection{Nonrenormalization of $\lambda$ and $\lambda_1$}
Here, we show that at the one-loop order there are no relevant corrections to $\lambda$ and $\lambda_1$.
\begin{figure}[b]
\centering
\includegraphics[width=0.44\textwidth]{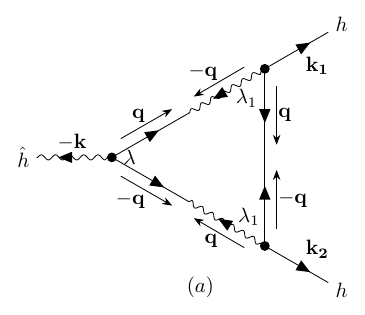}\hspace{0.5in}
\includegraphics[width=0.44\textwidth]{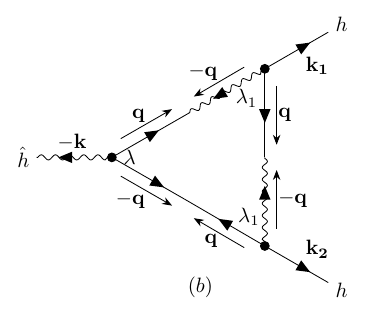}
\caption{One-loop Feynman diagrams giving the  corrections to  $\lambda$ and $\lambda_1$.}
\label{lam_diag}
\end{figure}
Fig.~\ref{lam_diag}(a) has a symmetry factor 24 and includes corrections to both $\lambda$ and $\lambda_1$. Contribution from this diagram is
\begin{align}
&-12D\lambda\lambda_1^2(k^2k_{1i}k_{2n})\int_{{\bf q},\Omega}\frac{q^6q_iq_n|{\bf q}|^{-y}}{(\Omega-i\nu q^4)^2(\Omega+i\nu q^4)^2} \nonumber \\ &=-3D\lambda\lambda_1^2(k^2k_{1i}k_{2n})\int_{{\bf q}}\frac{q_iq_n|{\bf q}|^{-y}}{\nu^3 q^6}.\label{lam_corr_1}
\end{align}
Similarly, Fig.~\ref{lam_diag}(b) has a symmetry factor $48$ and contributes
\begin{align}
&-24D\lambda\lambda_1^2(k^2k_{1i}k_{2n})\int_{{\bf q},\Omega}\frac{q^6q_iq_n|{\bf q}|^{-y}}{(\Omega-i\nu q^4)^3(\Omega+i\nu q^4)} \nonumber \\ &=3D\lambda\lambda_1^2(k^2k_{1i}k_{2n})\int_{{\bf q}}\frac{q_iq_n|{\bf q}|^{-y}}{\nu^3 q^6}.\label{lam_corr_2}
\end{align}
Equations~(\ref{lam_corr_1}) and (\ref{lam_corr_2}), when added, gives zero contribution to the corrections of $\lambda$ and $\lambda_1$. 

\subsection{RG recursion relations}

In the above, we have integrated out the dynamical fields with wavevectors $|{\bf k}|$ in the range ($\Lambda/b \leq |{\bf k}| \leq \Lambda$) in the action functional (\ref{action}). Next, we rescale space, time and fields to raise the  wavevector cutoff of $|{\bf k}|$ back to $\Lambda$. We rescale
 ${\bf k} \to {\bf k}^{'}=b{\bf k} \implies {\bf x} \to {\bf x}^{'}=\frac{{\bf x}}{b}$ similarly, $\omega \to \omega^{'}=b^z\omega \implies t{'}=\frac{t}{b^z}$. Furthermore, the fields are rescaled as
\begin{align*}
    &h^<({\bf k},\omega)=\xi h({\bf k}^{'},\omega^{'}),\\
     &\hat h^<({\bf k},\omega)=\hat \xi \hat h({\bf k}^{'},\omega^{'}).
\end{align*}
Using these we find that, model parameters scale as
\begin{align*}
&\nu'=\nu^<b^{-d-4-z}\xi\hat\xi,\\
&D'=D^<b^{y-d-z}\hat\xi^{2},\\
   &(\lambda',\lambda_1')=(\lambda^<,\lambda_1^<) b^{-2d-4-2z}\hat\xi\xi^2.
\end{align*}
 Non-renormalization of $D$, $\lambda$, and $\lambda_1$ yields $(D^<,\lambda^<,\lambda_1^<)\equiv(D,\lambda,\lambda_1)$. Also, $h^<({\bf x},t)$ scale as $b^{-d-z}\xi h({\bf x}^{'},t^{'})$. Defining $\xi_R=b^{-d-z}\xi=b^\chi$ and imposing the condition that coefficient of $\int_{{\bf x},t}\hat{h}\partial_th$ remains unity under rescaling, we get model parameters rescale as
\begin{align*}
&\nu'=\nu^<b^{z-4},\\
&D'=Db^{z+y-d-2\chi},\\
&(\lambda',\lambda_1')=(\lambda,\lambda_1)b^{ \chi+z-4}.
\end{align*}
Here, $\chi$ is the roughness exponent. Now we set $b=e^{\delta l}\approx 1+\delta l$ for small $\delta l$, and define a dimensionless coupling constant by $g=k_d\frac{\lambda^2D}{\nu^3}$ to obtain the flow equations
\begin{align}
&\frac{d\nu}{dl} = \nu\biggl[z-4+g[\frac{6+y-d}{4d}+\gamma\biggl(\frac{3+d+y}{2d}+\frac{3(6+y)}{2d(d+2)}\biggl)+\gamma^2\biggl(\frac{3(4+y)}{d(d+2)}+\frac{1}{d}\biggl)]\biggl],\\
&\frac{dD}{dl}=D(z+y-d-2\chi),\\
&\frac{d(\lambda,\lambda_1)}{dl}=(\lambda,\lambda_1)(\chi+z-4),\\
&\frac{dg}{dl}=g\biggl[4+y-d-3g\biggl[\frac{6+y-d}{4d}+\gamma\biggl(\frac{3+d+y}{2d}+\frac{3(6+y)}{2d(d+2)}\biggl)+\gamma^2\biggl(\frac{3(4+y)}{d(d+2)}+\frac{1}{d}\biggl)\biggl]\biggl].
\end{align}
See Eqs.~(5)-(8) of the main text.

\section{MCT Calculations}

We start with the general definitions of the correlation function $C({ {\bf k}},\omega)$ and response function $G({ {\bf k}},\omega)$:
\begin{align}
&\delta^{(d)}({\bf k+k^{'}})\delta(\omega+\omega^{'})G({ {\bf k}},\omega)=\biggl\langle\frac{\delta h({\bf k},\omega)}{\delta \eta({\bf k^{'}},\omega^{'})}\biggr\rangle\\
&\delta^{(d)}({\bf k+k^{'}})\delta(\omega+\omega^{'})C({ {\bf k}},\omega)=\langle h({\bf k},\omega)h({\bf k^{'}},\omega^{'})\rangle.
\end{align}
The response and correlation functions are assumed to have the following scaling form in the long time large length scale limit:
\begin{align}
&G({\bf k},\omega)=k^{-z}g\biggl(\frac{\omega}{k^z}\biggr)\\
&C({\bf k},\omega)=k^{-y-2z}f\biggl(\frac{\omega}{k^z}\biggr),
\end{align}
where $z$ is the dynamic exponent and $\chi$ is the roughness exponent; see also the main text for their definitions. The Dyson's equation for the self energy ($\Sigma$) is given by
\begin{align}
G^{-1}({\bf k},\omega)=-i\omega+\Sigma({\bf k},\omega).
\end{align}
The zero frequency self energy or the relaxation rate has the form
\begin{align}
\Sigma({\bf k},0)=\Gamma k^z. \label{selfeng_1}
\end{align}
Here $\Gamma>0$ is the effective damping coefficient. The correlation function in Lorentzian approximation can be written as
\begin{align}
C({\bf k},\omega)=\frac{2D |{\bf k}|^{-y}}{\omega^2+\Gamma^2k^{2z}},
\end{align}
giving the zero-frequency correlation function as
\begin{align}
C({\bf k},0)=\frac{2D}{\Gamma^2} |{\bf k}|^{-y-2z}.\label{correl_1}
\end{align}
\begin{figure}[htb]
\includegraphics[width=0.44\textwidth]{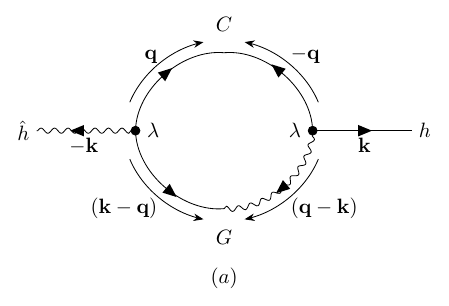}
\includegraphics[width=0.44\textwidth]{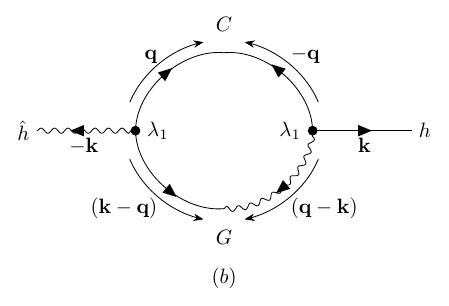}\\
\includegraphics[width=0.44\textwidth]{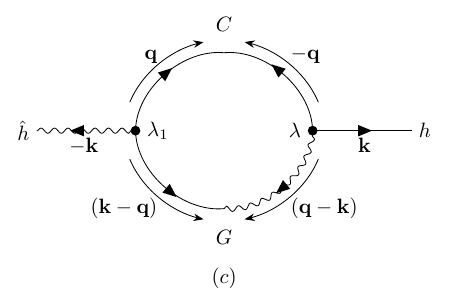}
\includegraphics[width=0.44\textwidth]{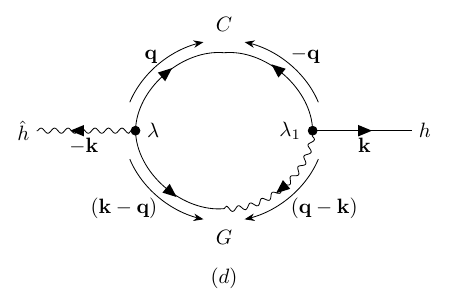}
\caption{ One-loop Feynman diagrams that contribute to the zero frequency self energy in the MCT.}
\label{self_eng_diag}
\end{figure}
Fig.~\ref{self_eng_diag}(a) shows the one-loop diagrammatic correction to the self energy at ${\cal O}(\lambda^2)$ that comes with a symmetry factor of $8$ and contributes
\begin{align}
&\lambda^2\int_{{\bf q},\Omega}k^2{\bf q\cdot(k-q)}(q-k)^2{\bf (q\cdot k)}C({\bf q}, \Omega)G({\bf k-q},-\Omega)
=\lambda^2k^2k_i\int_{{\bf q},\Omega}\frac{{\bf q \cdot (k-q)}(q-k)^2q_i (2D)q^{-y}}{(\Omega^2+\Gamma^2q^{2z})(i\Omega+\Gamma{\bf |k-q|}^z)}.
\end{align}
After performing the $\Omega$-integral, above integral becomes
\begin{align}
\frac{\lambda^2 D}{\Gamma^2}k^2k_i\int_{{\bf q}}\frac{q_j(k-q)_j(q-k)^2q_i q^{-y-z}}{(q^z+|{\bf k-q}|^z)}.
\end{align}
After symmetrization i.e. ${\bf q}\rightarrow{\bf q}+\frac{\bf k}{2}$, above integral simplifies to two integrals
\begin{align}
&\mathcal {I}_{11}=-k_d\frac{\lambda^2 D}{4\Gamma^2}k^4\int_k^{\Lambda}q^{-y-2z+4+d-1}dq=-k_d\frac{\lambda^2 D}{4\Gamma^2}k^4\biggl[\frac{k^{4+d-y-2z}-\Lambda^{4+d-y-2z}}{y+2z-4-d}\biggl],\label{i11}\\
&\mathcal {I}_{12}=\frac{\lambda^2 D}{2\Gamma^2}k^2k_ik_m(2+y+z)\int_{\bf q}\frac{q_iq_m}{2q^z}q^{-y-z+2}=k_d\frac{\lambda^2 D}{4\Gamma^2}\biggl(\frac{2+y+z}{d}\biggl)k^4\biggl[\frac{k^{4+d-y-2z}-\Lambda^{4+d-y-2z}}{y+2z-4-d}\biggl].\label{i12}
\end{align}
Adding the above contributions, we get the ${\cal O}(\lambda^2)$ correction to the zero frequency self energy
\begin{align}
k_d\frac{\lambda^2 D}{4\Gamma^2}\biggl(\frac{2+y+z-d}{d}\biggl)k^4\biggl[\frac{k^{4+d-y-2z}-\Lambda^{4+d-y-2z}}{y+2z-4-d}\biggl].\label{s_e_1}
\end{align}


Fig.~\ref{self_eng_diag}(b) shows the one-loop contribution to the self energy at ${\cal O}(\lambda_1^2)$ with a symmetry factor of $8$ and contributes
\begin{align}
4\lambda_1^2\int_{{\bf q},\Omega}k_ik_jq_i(k-q)_j(q-k)_m(q-k)_nq_mk_n \frac{2D q^{-y}}{(\Omega^2+\Gamma^2q^{2z})({i\Omega+\Gamma{\bf |k-q|}^z})}.
\end{align}
After performing the $\Omega$-integral,  we get
\begin{align}
\frac{2D\lambda_1^2}{\Gamma^2}k_ik_jk_n\int_{{\bf q}}\frac{q_i(k-q)_j(q-k)_m(q-k)_nq_mq^{-y-z}}{q^z}.
\end{align}
Then shifting ${\bf q}\rightarrow{\bf q}+\frac{\bf k}{2}$ and extracting the ${\cal O}(k)$ contribution from the numerator, above integral reduces to two separate integrals
\begin{align}
&\mathcal {I}_{13}=\frac{2\lambda_1^2 D}{\Gamma^2}k_ik_jk_mk_n\Bigl(\frac{y+z}{2}\Bigl)\int_{\bf q}\frac{q_iq_jq_mq_nq^{-y-z}}{q^z}=k_d\frac{3\lambda_1^2 D}{\Gamma^2}\frac{(y+z)}{d(d+2)}k^4\biggl[\frac{k^{4+d-y-2z}-\Lambda^{4+d-y-2z}}{y+2z-4-d}\biggl],\label{i13}\\
&\mathcal {I}_{14}=\frac{\lambda_1^2 D}{\Gamma^2}k^2k_ik_j\int_{\bf q}\frac{q_iq_jq^2q^{-y-z}}{q^z}=k_d\frac{\lambda_1^2 D}{\Gamma^2}\frac{1}{d}k^4\biggl[\frac{k^{4+d-y-2z}-\Lambda^{4+d-y-2z}}{y+2z-4-d}\biggl].\label{i14}
\end{align}
Adding $\mathcal {I}_{13}$ and $\mathcal {I}_{14}$ we get contribution to zero frequency self energy at ${\cal O}(\lambda_1^2)$:
\begin{align}
k_d\frac{\lambda_1^2 D}{\Gamma^2}\biggl(\frac{1}{d}+\frac{3(y+z)}{d(d+2)}\biggl)k^4\biggl[\frac{k^{4+d-y-2z}-\Lambda^{4+d-y-2z}}{y+2z-4-d}\biggl].\label{s_e_2}
\end{align}
Fig.~\ref{self_eng_diag}(c) gives one of the two one-loop corrections to the  self energy at  ${\cal O}(\lambda\lambda_1)$, with a symmetry factor of $4$ and contributes
\begin{align}
2\lambda \lambda_1\int_{{\bf q},\Omega}k_ik_jq_i(k-q)_j(q-k)^2({\bf q\cdot k}) \frac{2D q^{-y}}{(\Omega^2+\Gamma^2q^{2z})(i\Omega+\Gamma{ \bf |k-q|}^z)}.
\end{align}
After performing the $\Omega$-integral we get
\begin{align}
\frac{\lambda \lambda_1 D}{\Gamma^2}k_ik_jk_m\int_{{\bf q}}\frac{q_i(k-q)_j(q-k)^2q_mq^{-y-z}}{q^z}.
\end{align}
Again shifting ${\bf q}\rightarrow{\bf q}+\frac{\bf k}{2}$ and extracting the ${\cal O}(k)$ contribution from the numerator, above integral reduces to two separate integrals
\begin{align}
&\mathcal {I}_{15}=\frac{\lambda\lambda_1 D}{\Gamma^2}k_ik_jk_mk_n\int_{\bf q}\frac{q_iq_jq_mq_nq^{-y-z}}{q^z}=k_d\frac{\lambda\lambda_1 D}{\Gamma^2}\frac{3(2+y+z)}{2d(d+2)}k^4\biggl[\frac{k^{4+d-y-2z}-\Lambda^{4+d-y-2z}}{y+2z-4-d}\biggl],\label{i15}\\
&\mathcal {I}_{16}=-\frac{\lambda\lambda_1 D}{2\Gamma^2}k_ik_jk^2\int_{\bf q}\frac{q_iq_jq^{-y-z+2}}{q^z}=-k_d\frac{\lambda\lambda_1 D}{2\Gamma^2}\frac{1}{d}k^4\biggl[\frac{k^{4+d-y-2z}-\Lambda^{4+d-y-2z}}{y+2z-4-d}\biggl].\label{i16}
\end{align}
Finally, Fig.~\ref{self_eng_diag}(d) gives the second one-loop correction to the self energy that is also ${\cal O}(\lambda\lambda_1)$, but is distinct from the one in Fig.~\ref{self_eng_diag}(c):
\begin{align}
2\lambda \lambda_1k^2k_j\int_{{\bf q},\Omega} q_m(k-q)_m(q-k)_i(q-k)_jq_i\frac{2Dq^{-y}}{(\Omega^2+\Gamma^2q^{2z})(i\Omega+\Gamma { \bf |k-q|}^z)}.
\end{align}
After performing the $\Omega$-integral, above integral becomes
\begin{align}
\frac{D\lambda \lambda_1}{\Gamma^2}k^2k_j\int_{{\bf q}}\frac{q_m(k-q)_m(q-k)_i(q-k)_jq_iq^{-y-z}}{q^z}.
\end{align}
After shifting ${\bf q}\rightarrow{\bf q}+\frac{\bf k}{2}$, above integral reduces to two integrals
\begin{align}
&\mathcal {I}_{17}=\frac{\lambda\lambda_1 D}{\Gamma^2}\Bigl(\frac{y+z}{2}\Bigl)k^2k_jk_m\int_{\bf q}\frac{q_jq_mq^{2-y-z}}{q^z}=k_d\frac{\lambda\lambda_1 D}{\Gamma^2}\frac{(y+z)}{2d}k^4\biggl[\frac{k^{4+d-y-2z}-\Lambda^{4+d-y-2z}}{y+2z-4-d}\biggl],\label{i17}\\
&\mathcal {I}_{18}=\frac{\lambda\lambda_1 D}{2\Gamma^2}k^4\int_{\bf q}\frac{q^{4-y-z}}{q^z}=k_d\frac{\lambda\lambda_1 D}{2\Gamma^2}k^4\biggl[\frac{k^{4+d-y-2z}-\Lambda^{4+d-y-2z}}{y+2z-4-d}\biggl].\label{i18}
\end{align}
Adding integrals $(\mathcal {I}_{15})$-$(\mathcal {I}_{18})$ we get the total ${\cal O}(\lambda\lambda_1)$ contribution to  self energy at zero-frequency
\begin{align}
k_d\frac{\lambda\lambda_1 D}{\Gamma^2}\biggl(\frac{y+z+d-1}{2d}+\frac{3(2+y+z)}{2d(d+2)}\biggl)k^4\biggl[\frac{k^{4+d-y-2z}-\Lambda^{4+d-y-2z}}{y+2z-4-d}\biggl].\label{s_e_3}
\end{align}
Adding Eqs.~(\ref{s_e_1}), (\ref{s_e_2}) and (\ref{s_e_3}) we obtain total contribution to zero frequency self energy,
\begin{align*}
&k_d\frac{\lambda^2 D}{\Gamma^2}\biggl[\frac{2+y+z-d}{4d}+\gamma\biggl(\frac{y+z+d-1}{2d}+\frac{3(2+y+z)}{2d(d+2)}\biggl)+\gamma^2\biggl(\frac{1}{d}+\frac{3(y+z)}{d(d+2)}\biggl)\biggl]k^4\biggl[\frac{k^{4+d-y-2z}-\Lambda^{4+d-y-2z}}{y+2z-4-d}\biggl],\nonumber\\&=k_d\frac{\lambda^2 D}{\Gamma^2}\tilde\Delta_m (\gamma,y,z)k^4\biggl[\frac{k^{4+d-y-2z}-\Lambda^{4+d-y-2z}}{y+2z-4-d}\biggl]
\end{align*}
Here,
\begin{align}
\tilde\Delta_m (\gamma,y,z)=\biggl[\frac{2+y+z-d}{4d}+\gamma\biggl(\frac{y+z+d-1}{2d}+\frac{3(2+y+z)}{2d(d+2)}\biggl)+\gamma^2\biggl(\frac{1}{d}+\frac{3(y+z)}{d(d+2)}\biggl)\biggl].
\end{align}
At the critical dimension obtained from the RG analysis, i.e., $d = d_c = 4 + y$, equating the powers of $k$ in the above expression with $\Gamma k^z$ yields the dynamic exponent $z = 4$. Substituting $z=4$ in above expression we obtain correction to self energy at critical dimension,
\begin{align}
&k_d\frac{\lambda^2 D}{\Gamma^2}\ln(\Lambda/k)\frac{1}{2(4+y)}\biggl[1+\gamma(10+2y)+\gamma^2\biggl(\frac{8y+36}{6+y}\biggl)\biggl],\nonumber\\
&=k_d\frac{\lambda^2 D}{\Gamma^2}\ln(\Lambda/k)\Delta_m(\gamma,y).
\end{align}
Here,
\begin{align}
\Delta_m(\gamma,y)=\frac{1}{2(4+y)}\biggl[1+\gamma(10+2y)+\gamma^2\biggl(\frac{8y+36}{6+y}\biggl)\biggl].
\end{align}
as given in the main text.

\section{Numerical Details}

\begin{figure}[t]
\centering
\includegraphics[width=0.8\textwidth]{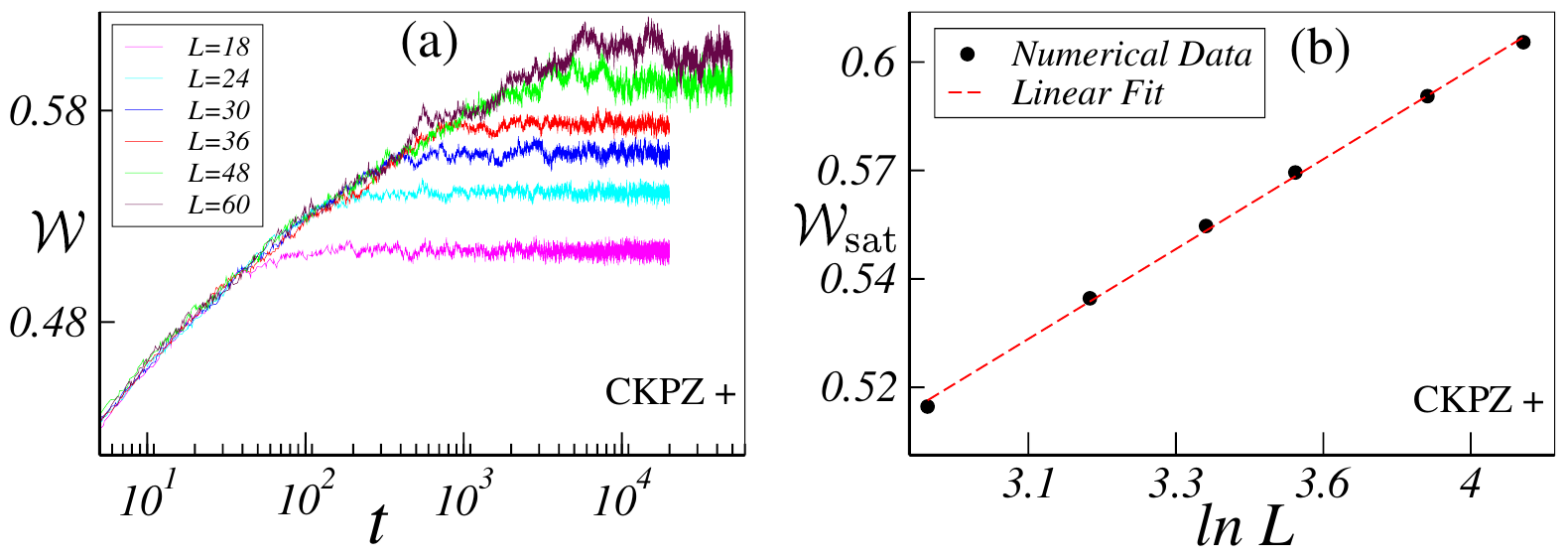}
\caption{Log-log plots of (a) width (${\mathcal W}$) versus time ($t$) for the CKPZ+ model for different system sizes. Width saturates in the stable region. (b) ${\mathcal W}_{\text{sat}}$ versus $\ln L$ for the CKPZ+ model. The red dashed line represent linear fit to the numerical data displayed on a log-log scale, showing logarithmic scaling behavior. These results are obtained for parameters $\bar\nu=0, \nu=1$, $D=1$, $\nu_r=0.2$, $\lambda=2$, $\lambda_1 = 1$, and time step $(\delta t) = 10^{-3}$, {using Euler scheme for time integration.}}
\label{plot_ckpz_plus}
\end{figure}
To suppress numerical instabilities, we incorporate a higher-order regularization term
$\nu_r \nabla^6 h$ ($\nu_r>0$) into Eq.~(1) of the main text. The evolution equation for the
height field then reads
\begin{equation}
\partial_t h = \bar\nu\nabla^2h-\nu \nabla^4 h
-\frac{\lambda}{2} \nabla^2 (\boldsymbol{\nabla} h)^2
-\lambda_1 \nabla_i \nabla_j (\nabla_i h \, \nabla_j h)
+ \nu_r \nabla^6 h
+ \eta . \label{mbeckpz_reg}
\end{equation}
We follow Ref.~\cite{ckpz_plus} and employ a pseudo-spectral method with the $2/3$ dealiasing rule, combined with an {Euler and a Heun time-integration scheme~\cite{euler}}, to simulate both MBE+ and CKPZ+ models in $d=2$. Unless stated otherwise, the parameters are fixed at $\bar\nu=0, \nu=1$, $D=1$ and $\lambda=2$. Simulations are performed on a square lattice with unit lattice spacing $a_0=1$.

We first focus on the stable regime. For $\lambda_1 = 1$, corresponding to $\gamma = 0.5$, which lies in the stable region of Fig.~2 of the main text, simulations are run up to total time $T=20000$ for system sizes $L \leq 36$ and up to $T=50000$ for larger systems ($L > 36$), using a time
step $\delta t = 10^{-3}$. Fig.~\ref{plot_ckpz_plus}(a) shows the temporal evolution of the surface
width $\mathcal{W}(t)$ for various $L$, averaged over $R$ realizations
($R = 1000$ for $L=18$ and $R=50$ for $L=60$). The width saturates after a finite time, as expected. Fig.~\ref{plot_ckpz_plus}(b) presents the scaling of the saturated width $\mathcal{W}_\mathrm{sat}$
with $\ln L$ on a log-log plot. The data exhibit clear logarithmic scaling, consistent with
theoretical predictions. Results for the MBE+ model are discussed in the main text.

\begin{figure}[t]
\centering
\includegraphics[width=1.0\textwidth]{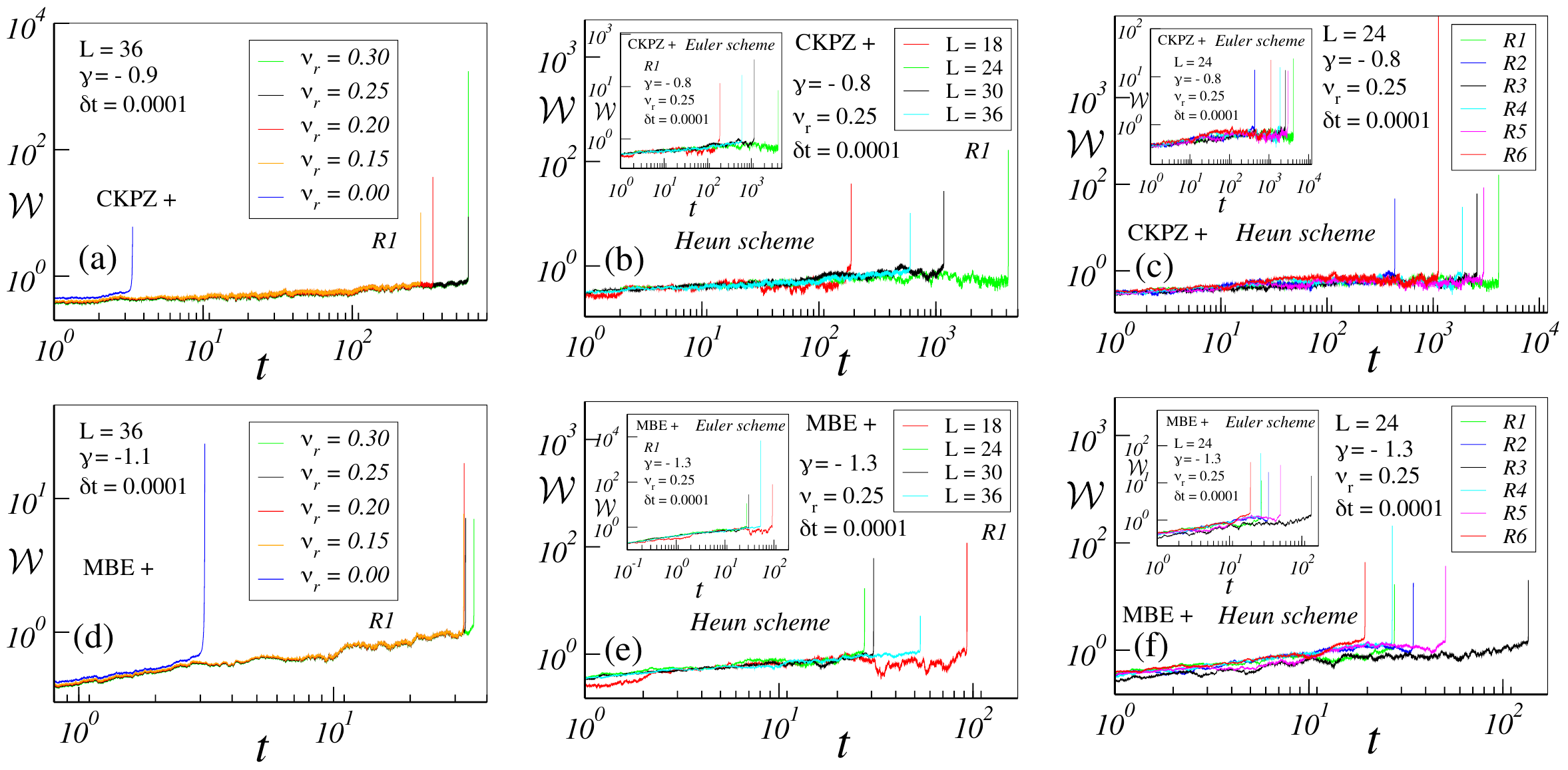}
\caption{ Time evolution of $\mathcal W$ demonstrating the robustness of the runaway behavior in the unstable regime.
(a,d) Results for different values of the regularization parameter $\nu_r$, showing that the runaway persists for all $\nu_r$, with only a shift in the onset time, for (CKPZ+, MBE+) model, obtained using Heun scheme for time interation.
(b,e) Comparison between Euler and Heun schemes, for different system sizes, indicating that the runaway behavior is similar in both schemes.
(c,f) Comparison between Euler and Heun schemes, for multiple independent noise realizations $(R1$-$R6)$, for a fixed system size $L=24$, again confirming similar growth behavior. All these observations suggest that the instability is independent of the chosen regularization and also independent of specific algorithmic set up. Specific models or integration schemes or numerical parameters are mentioned in the panels.
}
\label{plot_heun}
\end{figure}

To explore the dynamics in the unstable regime, we fix the time step at $\delta t = 10^{-4}$ and
set $\lambda = 2$, while varying $\lambda_1$ to tune the parameter $\gamma$. {To establish that the observed runaway behavior in the unstable regime is not a consequence of the chosen regularization or numerical algorithmic set up, we have reexamined Eq.~(\ref{mbeckpz_reg}) using a higher-order Heun scheme~\cite{euler} for time integration. In addition, the strength of the higher-order regularizing term $\nu_r$ has been varied over a wide range, including the limiting case $\nu_r = 0$. As shown in Fig.~\ref{plot_heun}(a,d) for the CKPZ+ and MBE+ models (for a representative realization $R1$), the runaway behavior persists for all values of $\nu_r$. Although the onset time shifts with increasing $\nu_r$, the eventual divergence of the width remains unchanged, indicating that the instability is not an artifact of the regularization.

To further rule out possible numerical artifacts associated with the numerical algorithmic set up, we have compared results obtained using both the Euler and Heun schemes. As illustrated in Fig.~\ref{plot_heun}(b,e), for both CKPZ+ and MBE+ models in the unstable regime (at fixed $\gamma$), the two schemes produce qualitatively identical runaway behavior across different system sizes $L$.

We have also examined the effect of statistical fluctuations by considering multiple independent noise realizations. Figure~\ref{plot_heun}(c,f) shows the time evolution of the width for several realizations at a fixed system size $L=24$. All realizations display similar runaway growth, demonstrating that the instability is not driven by rare events.

Taken together, the insensitivity of the runaway behavior to variations in regularization, numerical algorithmic set up, system size, and noise realizations provides strong evidence that the observed instability is intrinsic to the dynamics.}

For completeness, we summarize below the numerical protocol used in the simulations:
\begin{itemize}
    \item \textbf{Integration scheme:} Pseudo-spectral method with the $2/3$ dealiasing rule.
    \item \textbf{Time stepping:} Explicit Euler and Heun scheme~\cite{euler}.
    \item \textbf{Fixed parameters:} $\bar\nu=0, \nu=1$, $D=1$ and $\lambda=2$.
    \item \textbf{Lattice:} Square grid with unit spacing $a_0=1$.
    \item \textbf{Simulation time:} $T=20000$ for $L \leq 36$, and $T=50000$ for $L > 36$ in stable region.
    \item \textbf{Time step:} $\delta t = 10^{-3}\, (10^{-4})$ in the stable (unstable) regime.
\end{itemize}

\end{document}